\documentclass[conference]{IEEEtran}
\IEEEoverridecommandlockouts
\usepackage{cite}
\usepackage{amsmath,amssymb,amsfonts}
\usepackage{algorithmic}
\usepackage{graphicx}
\usepackage{textcomp}
\usepackage{xcolor}
\allowdisplaybreaks
\def\BibTeX{{\rm B\kern-.05em{\sc i\kern-.025em b}\kern-.08em
    T\kern-.1667em\lower.7ex\hbox{E}\kern-.125emX}}
\begin{document}

\title{An Integration and Operation Framework of Geothermal Heat Pumps in Distribution Networks
\thanks{This work was supported by Tsinghua-Berkeley Shenzhen Institute Research Start-Up Funding and Shenzhen Science and Technology Program (Grant No. KQTD20170810150821146). L. Liang, X. Zhang and H. Sun are with Tsinghua-Berkeley Shenzhen Institute, Tsinghua University, Shenzhen 518055, Guangdong, China (email:liangl18@mails.tsinghua.edu.cn, xuanzhang@sz.tsinghua.edu.cn, shb@mail.tsinghua.edu.cn). \textit{Corresponding Author: Xuan Zhang.}}}
\author{Lei Liang, Xuan Zhang* and Hongbin Sun \textit{Fellow}, IEEE}
\maketitle

\begin{abstract}
The application of the energy-efficient thermal and energy management mechanism in Geothermal Heat Pumps (GHPs) is indispensable to reduce the overall energy consumption and carbon emission across the building sector. Besides, in the Virtual Power Plant (VPP) system, the demand response of clustered GHP systems can improve the operating flexibility of the power grid. This paper presents an integration and operation framework of GHPs in the distribution network, applying a layered communication and  optimization method to coordinate multiple clustered GHPs in a community. In the proposed hierarchical operation scheme, the operator of regional GHPs collects the thermal zone information and the disturbance prediction of buildings in a short time granularity, predicts the energy demand, and transmits the information to an aggregator. Using a novel linearized optimal power flow model, the aggregator coordinates and aggregates load resources of GHP systems in the distribution network. In this way, GHP systems with thermal and energy management mechanisms can be applied to achieve the demand response in the VPP and offer more energy flexibility to the community.

\end{abstract}


\section{Introduction}
In a Virtual Power Plant (VPP), the effective demand-side management of Distributed Energy Resources (DERs) and responsive loads can improve the operational flexibility of distribution networks~\cite{b1}. Additionally, accounting for about 50\% of building energy consumption globally~\cite{BC1}, the Heating, Ventilation, and Air-Conditioning (HVAC) system, a kind of DER, has a huge potential in energy saving. Load shedding of large-scale HVAC systems during peak demand not only enables the demand response of the VPP system to provide more energy flexibility for communities but also makes a significant contribution to reducing the overall global energy consumption and carbon emissions~\cite{HVAC1}. Therefore, the community energy management of HVAC systems in distribution networks is attracting growing attention.

Geothermal energy is a distributed renewable energy. The typical application of geothermal energy is the Geothermal Heat Pump (GHP) system, a novel energy-efficient HVAC system, using low-temperature resources of shallow soil and groundwater around 5$\sim$30℃ to indirectly heat and cool buildings~\cite{self}. The GHP system is becoming more and more popular for its excellent performance in reducing carbon emissions and meeting the growing energy demand~\cite{attention}. Researchers have proposed various methods to realize the optimization and control of GHP systems, e.g., classical or traditional control methods such as ON/OFF control~\cite{onoff} and PID control~\cite{pid}, and other advanced control technologies such as model predictive control~\cite{mpc}, neural network-based control~\cite{neural}, hybrid or fusion control~\cite{fusion} etc. Most of those methods are only intended to ensure that a single GHP system operates efficiently and reliably in one building. On the other hand, in the multi-buildings community, it is necessary to establish an easily implemented and extensible hierarchical optimization and control scheme~\cite{dsm}, expecting to realize the active and rapid response of the large-scale GHP systems in the VPP network. 

Some researches have focused on the hierarchical demand response framework \cite{GC1,GC3,GC4}. In reference \cite{GC1}, a method using distributed model predictive control was proposed, and an integrated hierarchical framework for Coefficient of Performance (COP) and cost optimization was established. The work \cite{GC3} proposed three demand response control algorithms based on real-time/previous/forecast hourly electricity price to obtain the cost-optimal solution via the GHP system. In reference \cite{GC4}, a feedforward artificial neural network algorithm was presented, for the short-term load prediction of decomposing sites to realize demand response. However, the proposed framework in most literature requires a large amount of sensing, communication, and computation, which is only suitable for a small-scale domestic energy network instead of a large-scale one with multiple GHPs. In addition, the proposed thermal models of the GHP system are mostly linear, so it is difficult to capture the temperature dynamics in different thermal zones.

\textbf{Contributions}: This paper proposes an integration and operation framework, using hierarchical optimization methods to coordinate multiple clustered GHPs. Based on the higher-order thermal dynamic models for radiator heating/cooling, each operator of clustered GHPs collects thermal zone information and the disturbance prediction of buildings, predicts the energy demand in a short time granularity, and transmits the information to an aggregator. Creatively, we utilize a novel fully linear Optimal Power Flow (OPF) model to quantify the active and reactive power of multiple clustered GHPs, to coordinate the load resources and realize demand response. 

\section{Problem Formulation}
\subsection{The GHP System}
\begin{figure}[!t]
\centering
\includegraphics[width=0.41\textwidth,trim=40 60 20 20,clip]{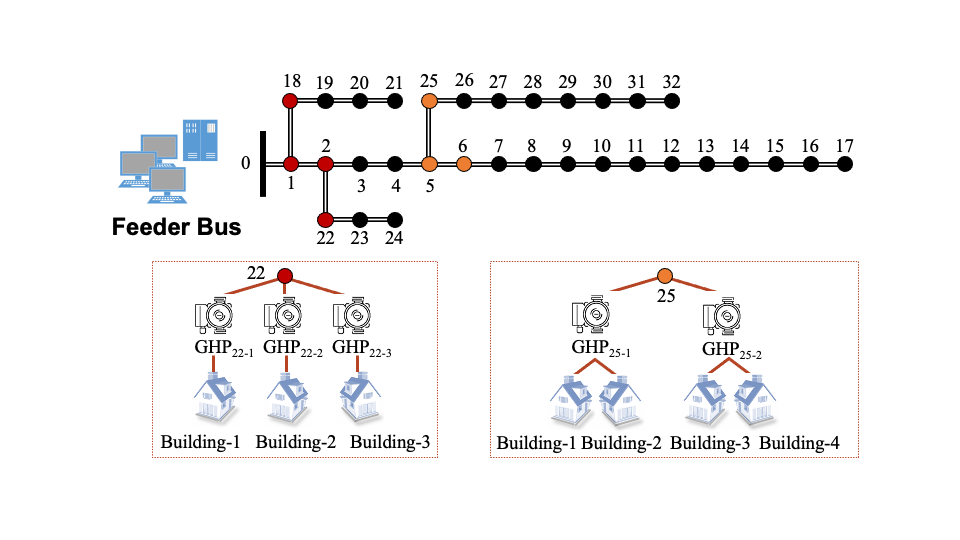}
\caption{Multiple clustered GHPs are integrated in the distribution network.}
\label{der}
\vspace{-0.5cm}
\end{figure}

The integration and operation framework for multiple clustered GHPs being integrated into a distribution network is shown in Figure~\ref{der}. In this subsection, the operating principle and thermodynamic models of the GHP systems in community-level buildings are introduced. Typically, a GHP system primarily consists of three subsystems: the underground heat exchange subsystem, the heat pump subsystem, and the heat distribution subsystem. Pipes with a hydraulic circuit in the heat exchange subsystem are employed to extract thermal energy from the ground or store heat for heating in winter and cooling in summer in the heat pump subsystem. In the heat distribution subsystem, the GHP system realizes space heating/cooling via radiator systems or floor radiant heating systems through thermal convection of input and extraction of water in pipes with indoor and outdoor air. The heat transfer mechanism in this paper only considers the major heat conduction processes through the inner walls and windows, which mainly depends on the flow rate and the supply temperature of the heat pump.


For simplicity, we only focus on radiator heating. The potential process of temperature dynamic evolution in buildings is complex and uncertain, so the Resistance-Capacitance (RC) network model is usually used to simulate it. Reference~\cite{model1} compared the response between a high-dimensional model and a low-order model, and the results showed that the second-order model could reproduce the input-output behavior of a full-scale model (with 13 states) better. Consider clustered GHP systems in a certain district for radiator heating/cooling, denote $\mathcal{M}$ as the set of GHPs in a district for radiator heating/cooling, $\mathcal{K}_m,m\in\mathcal{M}$ as the set of the buildings in the communities with GHP $m$, and $\mathcal{R}_k,k\in\mathcal{K}_m$ as a set of rooms/zones. Each building is modeled as a connected undirected graph $(\mathcal{R}_k,\mathcal{E}_k)$, where $\mathcal{R}_{k}$ represents the nodes collection of rooms in building $ k\in\mathcal{K}=\cup_{m\in\mathcal{M}}\mathcal{K}_m$ and $\mathcal{E}_k\subseteq\mathcal{R}_{k}\times\mathcal{R}_{k}$ represents the edges collection. If room $i$ is adjacent to room $j$ in building $k$, there exists an edge $(i,j)$ in the collection $\mathcal{E}_k$. Denote $\mathcal{R}_{k}(i)$ as the collection of adjacent rooms for room $i$ in building $k$. A second-order RC model is introduced for describing the thermal models of zones:
\begin{subequations}\label{RC}
\begin{align}
C_{ki}\dot T_{ki}\!\!=&\frac{{T_k^o}\!\!-\!\!T_{ki}}{R_{ki}}\!+\!\!\!\!\!\!\sum_{j\in\mathcal{R}_k(i)}\!\!\!\!\frac{T_{kij}\!\!-\!\!T_{ki}}{R_{kij}}\!+\!\!\!\sum_{n_{ki}=1}^{N_{ki}}\!\!\!\!\frac{T_{n_{ki}}\!\!-\!T_{ki}}{R_{{ar}_{ki}}}\!+\!Q_{ki}\\
C_{kij}\dot T_{kij}&=\frac{T_{ki}-T_{kij}}{R_{kij}}+\frac{T_{kj}-T_{kij}}{R_{kij}}
\end{align}
\end{subequations}
while a high-order lumped element model is formulated for modeling the radiator with $N_{ki}$ elements.
\begin{equation}\label{PIPE}
C_{n_{ki}}\dot T_{n_{ki}}=\frac{T_{ki}-T_{n_{ki}}}{R_{{ar}_{ki}}}+c_{w}q_{ki}(T_{{n-1}_{ki}}-T_{n_{ki}})
\end{equation}
where $k\!\in\!\mathcal{K}_m$, $i\!\in\!\mathcal{R}_k$, $n_{ki}\!\!=\!\!1,2,\cdots,N_{ki}$ (${n\!-\!1}_{ki}\!=n_{ki}-\!1$), $c_{w}$ is the specific heat of the water, $q_{ki}$ is the water flow rate, $C_{\{\cdot\}}$ is the thermal capacitance, $R_{\{\cdot\}}$ is the thermal resistance, $R_{ar_{ki}}$ is the thermal resistance between radiator and indoor air, and $T_{\{\cdot\}}$ is the temperature. Note that (i) the radiator pipe is divided into $N_{ki}$ sections, and $N_{ki}$ varies among different rooms \cite{thesis}. As for the $n$th section, the entering water temperature is $T_{{n-1}_{ki}}$ ($T_{0_{ki}}=T_{s_m}$), and the leaving water temperature is $T_{{n}_{ki}}$, all surrounded by room temperature $T_{ki}$. The term $c_{w}q_{ki}(T_{{n-1}_{ki}}\!\!-\!\!T_{n_{ki}})$ is the heat transferred in the $n$th section. (ii) $T_{s_m}$ is a common factor for all buildings connected with the same heat pump $m\in\mathcal{M}$ in building set $\mathcal{K}_m$. (iii) ${T_k^o}$ is the outside temperature, and $Q_{ki}\!\!\ge\!\!0$ indicates the heat disturbances from external sources (e.g., user activity, solar radiation and device operation). 

\noindent \textit{Remark 1}. When the heat distribution subsystem works normally, (\ref{RC})-(\ref{PIPE}) asymptotically converges to an equilibrium point and the steady state is determined by disturbances $T_k^o$, $Q_{ki}$ and control inputs $q_{ki}$, $T_{s_{m}}$. Otherwise ($q_{ki}\!\!=\!\!0$), the equilibrium state for (\ref{RC})-(\ref{PIPE}) is only determined by $T_k^o$, $Q_{ki}$. Since $T_k^o$, $Q_{ki}$ change slowly when considering a short time granularity in the normal mode, we only need to design the dynamics of $q_{ki}$, $T_{s_{m}}$ to drive (\ref{RC})-(\ref{PIPE}) to the desired state set by users or administrators. See detailed derivation in \cite{zhang}.

\noindent \textit{Remark 2}. Let $B_{ar_{ki}}\!\!\!=\!\!\!1/R_{{ar}_{ki}}$, the energy consumption for heating/cooling (i.e., the heat loss in the water pipe), can be summed as $\sum_{n_{ki}=1}^{N_{ki}}B_{ar_{ki}}(T_{{n}_{ki}}\!\!-\!\!T_{ki})=c_{w}q_{ki}(T_{s_m}\!-T_{{N}_{ki}})$. By calculating, the terminal temperature is $T_{{N}_{ki}}\!=\!(\!1\!-\!(\frac{c_{w}q_{ki}}{c_{w}q_{ki}\!+\!B_{ar_{ki}}})^{N_{ki}})T_{ki}+\!(\frac{c_{w}q_{ki}}{c_{w}q_{ki}+B_{ar_{ki}}})^{N_{ki}}T_{s_m}$. In steady state, we use $u_{ki}=c_{w}q_{ki}[1\!-(\frac{c_wq_{ki}}{c_wq_{ki}+B_{ar_{ki}}})^{N_{ki}}](T_{s_m}-Z_{ki})$ to 
represent the energy consumption, where $Z_{\{\cdot\}}$ is the steady-state temperature of $T_{\{\cdot\}}$. In a heating mode, $T_{s_m}>T_{ki}, T_{s_m}>Z_{ki}, \forall m,$ $\forall k, \forall i$ hold; in a cooling mode, $T_{s_m}<T_{ki}, T_{s_m}<Z_{ki}, \forall m, \forall k, \forall i$ hold.

\subsection{Forecast of Short-Term Energy Demand}
In this subsection, we focus on the forecast of short-term energy demand based on the collected information of thermal zones and the disturbance prediction of buildings. According to (\ref{RC})-(\ref{PIPE}), the external dependent factors of the thermal dynamic models are disturbances $T_k^o,Q_{ki}$. In a short time granularity (e.g. 5 minutes), $T_k^o, Q_{ki}$ change slowly and can be regarded as constants. With stable external inputs $T_k^o, Q_{ki}$, the heat pump can rapidly drive the temperature in each zone to the steady-state under the operating constraints and the user comfort conditions, compared with the changing speed of disturbances. Many researches have focused on the field of forecasting in temperature and thermal disturbances. Since data prediction methods is not the focus of this paper, we only uses existing statistical methods to realize the prediction of $\tilde{T}_k^o,\tilde{Q}_{ki}$ as in~\cite{zhang}. 

With the short-term predicted value, when (\ref{RC}), (\ref{PIPE}) reach the steady state under a normal operation mode, we have:
\begin{subequations}\label{RCPIPE=0}
\begin{align}
\frac{\tilde{T}_k^o\!\!-\!\!Z_{ki}}{R_{ki}}\!+\!\!\!\!\!\!\sum_{j\in\mathcal{R}_k(i)}\!\!\!\!\frac{Z_{kij}\!\!-\!\!Z_{ki}}{R_{kij}}+\!\!\!\sum_{n_{ki}=1}^{N_{ki}}\!\!\!\frac{Z_{n_{ki}}\!\!-\!Z_{ki}}{R_{{ar}_{ki}}}\!+\!\tilde{Q}_{ki}\!=\!0\\
\frac{Z_{ki}-Z_{kij}}{R_{kij}}+\frac{Z_{kj}-Z_{kij}}{R_{kij}}=0\\
\frac{Z_{ki}-Z_{n_{ki}}}{R_{{ar}_{ki}}}+c_{w}q_{ki}(Z_{{n-1}_{ki}}-Z_{n_{ki}})=0.
\end{align}
\end{subequations}
With $u_{ki}\!\!=\!\!c_wq_{ki}[1\!-(\frac{c_wq_{ki}}{c_wq_{ki}+B_{ar_{ki}}})^{N_{ki}}](T_{s_m}\!\!-\!Z_{ki})$, then we have the steady-state equation:
\begin{equation*}\label{SSE}
\frac{{\tilde{T}_k^o}-Z_{ki}}{R_{ki}}\!+\!\!\!\!\!\sum_{j\in\mathcal{R}_k(i)}\!\!\!\!\frac{Z_{kij}-Z_{ki}}{2R_{kij}}+u_{ki}+\!\tilde{Q}_{ki}=0.\tag{4a}
\end{equation*}

Since the control inputs are $q_{ki}, T_{s_m}$, the operating constraints of GHP systems are $0\leq q_{ki}\leq \overline{q_{ki}}$ and $\underline{T_{s_m}}\leq T_{s_m} \leq \overline{T_{s_m}}$. According to a monotonicity analysis, $u_{ki}$ is a monotonic increasing function of the variable $q_{ki}$. Using $u_{ki}$ to substitute $q_{ki}$ in the operating constraints, we have the operating constraints of the GHP system:
\begin{equation*}\label{OC1}
0\leq\!\! u_{ki}\!\! \leq c_w\overline{q_{ki}}[1\!-\!(\frac{c_w\overline{q_{ki}}}{c_w\overline{q_{ki}}\!\!+\!\!B_{{ar}_{ki}}})^{N_{ki}}](T_{s_m}\!\!-\!\!Z_{ki})\tag{4b}
\end{equation*}

\noindent\begin{equation*}\label{OC2}
\underline{T_{s_m}}\leq T_{s_m} \leq \overline{T_{s_m}}.\tag{4c}
\end{equation*}Users can customize their desired temperature by the user comfort constraint:
\begin{equation*}\label{UC}
\underline{T_{ki}}\leq Z_{ki} \leq \overline{T_{ki}}.\tag{4d}
\end{equation*}Also the uncertainty of disturbances are constrained: 
\begin{equation*}\label{PC1}
\underline{T_k^o}\leq \tilde{T}_k^o \leq \overline{T_k^o}, \underline{Q_{ki}}\leq \tilde{Q}_{ki} \leq \overline{Q_{ki}}\tag{4e}
\end{equation*}
where [$\underline{T_{ki}}$,$\overline{T_{ki}}$] is the set range of the comfortable temperature, $\overline{q_{ki}}$ is the upper bound of the flow rate, [$\underline{T_{s_m}},\overline{T_{s_m}}$] is the range of the supply temperature, subject to the inherent attribute of heat pumps, [$\underline{T_k^o},\overline{T_k^o}$] and [$\underline{Q_{ki}},\overline{Q_{ki}}$] are ranges of the prediction uncertainty, determined by the prediction methods. 

Actually, Equation $(4)$ defines the stable state of a single GHP system. Since the active power consumption of the GHP system equals the energy consumption of its heat pump, we can predict the range and desired value of active power consumption $[\underline{p_m},\overline{p_m}], p_m^d\in[\underline{p_m},\overline{p_m}]$ for each GHP with Equation $(4)$.

\subsubsection{The upper bound} 
The produced heat from one GHP can be expressed as $\sum_{k\in\mathcal{K}}\sum_{i\in\mathcal{R}_k}u_{ki}, u_{ki}=c_wq_{ki}[1-(\frac{c_wq_{ki}}{c_wq_{ki}+B_{{ar}_{ki}}})^{N_{ki}}](T_{s_m}-Z_{ki}), i\in\mathcal{R}_k, k\in\mathcal{K}_m, m\in\mathcal{M}$. Obviously, finding the upper bound $\overline{p_m}$ is equivalent to solving the following optimization problem:
\begin{subequations}\label{upperbound-1}
\begin{gather}
\max_{Z_{ki},u_{ki},T_{s_m},\overline{u_m},\tilde{T}_k^o,\tilde{Q}_{ki}} \frac{\overline{u_m}}{b_m-a_mT_{s_m}}\tag{5a}\\
\text{s. t. } (4a)-(4e) \nonumber\\
\overline{u_m}=\sum_{k\in\mathcal{K}_m}\sum_{i\in\mathcal{R}_k}u_{ki}\tag{5b}
\end{gather}
\end{subequations}
where $i\in\mathcal{R}_k,k\in\mathcal{K}_m,m\in\mathcal{M}$. $-a_mT_{s_m}+b_m$ is the Coefficient of Performance (COP) of the GHP system, where $a_m,b_m$ are positive coefficients~\cite{GC1}. A high COP value indicates high efficiency. Since the above problem is nonconvex due to the non-convex objective function $(5a)$ and the non-convex constraint $(4b)$ with the decision variable $T_{s_m}$, we consider a modified version given by
\begin{subequations}\label{upperbound-2}
\begin{gather}
\max_{Z_{ki},u_{ki},\overline{u_m},\tilde{T}_k^o,\tilde{Q}_{ki}} \frac{\overline{u_m}}{b_m-a_m\overline{T_{s_m}}}\tag{6a}\\
\text{s. t. } (4a),(4c)-(4e),(5b)\nonumber\\
0\leq\!\! u_{ki}\!\! \leq c_w\overline{q_{ki}}[1\!-\!(\frac{c_w\overline{q_{ki}}}{c_w\overline{q_{ki}}\!\!+\!\!B_{{ar}_{ki}}})^{N_{ki}}](\overline{T_{s_m}}\!\!-\!\!Z_{ki}).\tag{6b}
\end{gather}
\end{subequations}\noindent\textbf{Theorem 1.} \textit{The optimal objective values of the optimization problems (5) and (6) are the same.}

\noindent\textit{Proof.} Since the decision variable $T_{s_m}$ has the limit [$\underline{T_{s_m}},\overline{T_{s_m}}$], the constraint set of problem (5) is a subset of that of (6), which means the value of the optimal objective function $\frac{\overline{u_m}}{b_m-a_mT_{s_m}}$ with decision variables $Z_{ki},u_{ki},T_{s_m},\overline{u_m},\tilde{T}_k^o,\tilde{Q}_{ki}$ in problem (5) is no more than that of $ \frac{\overline{u_m}}{b_m-a_mT_{s_m}}$ with decision variables $Z_{ki},u_{ki},\overline{u_m},\tilde{T}_k^o,\tilde{Q}_{ki},\underline{T_{s_m}}\leq T_{s_m}\leq \overline{T_{s_m}}$. By setting $T_{s_m}=\overline{T_{s_m}}$ in the objective function, when the problem is feasible, the optimal objective function value of (5) is the same as that of (6), achieved at $T_{s_m}=\overline{T_{s_m}}$.\hfill$\square$\par

The upper bound of active power consumption of one GHP is obtained as $\overline{p_m}=\overline{u_m}^*/(b_m-a_m\overline{T_{s_m}})$, where $\overline{u_m}^*$ is the optimal solution of problem (6). 

\subsubsection{The lower bound}
Similarly, finding the lower bound $\underline{p_m}$ is equivalent to solving the optimization problem (7):
\begin{subequations}\label{lowerbound}
\begin{gather}
\min_{Z_{ki},u_{ki},T_{s_m},\underline{u_m},\tilde{T}_k^o, \tilde{Q}_{ki}} \frac{\underline{u_m}}{b_m-a_mT_{s_m}}\tag{7a}\\
\text{s. t. } (4a)-(4e) \nonumber\\
\underline{u_m}=\sum_{k\in\mathcal{K}}\sum_{i\in\mathcal{R}_k}u_{ki}.\tag{7b}
\end{gather}
\end{subequations}
Since the objective of the optimization problem (7) is to search for the minimum, (7) cannot be convexified as the process of (5) to (6). However, (7) can be handled by scaling to get two approximate optimal values of the objective function (i.e., an aggressive lower bound and a conservative lower bound). 

\noindent \textit{An aggressive lower bound:}
\begin{subequations}\label{smallerlowerbound}
\begin{gather}
\min_{Z_{ki},u_{ki},T_{s_m},\underline{u_m},\tilde{T}_k^o, \tilde{Q}_{ki}} \frac{\underline{u_m}}{b_m-a_m\underline{T_{s_m}}}\tag{8a}\\
\text{s. t. } (4a)-(4e),(7b)\nonumber\\
0\leq u_{ki}\leq c_w\overline{q_{ki}}[1-(\frac{c_w\overline{q_{ki}}}{c_w\overline{q_{ki}}\!+\!\!B_{{ar}_{ki}}})^{N_{ki}}](\overline{T_{s_m}}\!\!-\!\!Z_{ki}).\tag{8b}
\end{gather}
\end{subequations}

\noindent \textit{A conservative lower bound:}
\begin{subequations}\label{largerlowerbound}
\begin{gather}
\min_{Z_{ki},u_{ki},T_{s_m},\underline{u_m},\tilde{T}_k^o, \tilde{Q}_{ki}} \frac{\underline{u_m}}{b_m-a_m\underline{T_{s_m}}}\tag{9a}\\
\text{s. t. } (4a)-(4e),(7b)\nonumber\\
0\leq u_{ki}\leq c_w\overline{q_{ki}}[1-(\frac{c_w\overline{q_{ki}}}{c_w\overline{q_{ki}}\!+\!\!B_{{ar}_{ki}}})^{N_{ki}}](\underline{T_{s_m}}\!\!-\!\!Z_{ki}).\tag{9b}
\end{gather}
\end{subequations}
Note that if problem (7) is feasible, problem (8) is feasible, while problem (9) may be infeasible. When solving the problems, we assume problems (7), (8), (9) are all feasible. Since the constraint set of problem (8) is a subset of that of (7), whose constraint set is a subset of that of (9), we have that the optimal objective function values $O_{(8)}^*, O_{(9)}^*, O_{(10)}^*$ satisfy $O_{(9)}^*\leq O_{(8)}^*\leq O_{(10)}^*$. 

The lower bound of active power consumption of one GHP is obtained as $\underline{p_m}=\underline{u_m}^*/(b_m-a_mT_{s_m}^*)$, where $\underline{u_m}^*,T_{s_m}^*$ are the optimal solution of the optimization problem (7)/(8)/(9), and $T_{s_m}^*=\underline{T_{s_m}}$ in (8)/(9). We could obtain the exact value of the lower bound of active power consumption of the system through solving problem (7), or the approximate values through problem (8)/(9).

\subsubsection{The desired consumption}
The desired active power consumption should be mainly related to the expectation of temperature and of the efficiency of heat pumps. Thus, we could obtain the desired value of active power consumption $p_m^d$ when the indoor temperature is close to the set point $T_{ki}^{set}$ and the energy efficiency of the GHP is maximized by maximizing the COP value (or minimizing the supply temperature $T_{s_m}$) concurrently:
\begin{subequations}\label{desired}
\begin{gather}
\min_{Z_{ki},u_{ki},T_{s_m}} \!\!\!\frac{\phi_a}{2}\!\sum_{k\in\mathcal{K}_m}\!\sum_{i\in\mathcal{R}_k}\!(Z_{{ki}}\!\!-\!\!T_{{ki}}^{set})^2\!+\!\frac{\phi_b}{2}(T_{s_m}\!\!\!-\!\!\underline{T_{s_m}})^2\tag{10a}\\
\text{s. t. } (4a)-(4e)\nonumber\\
\widetilde{T}_k^o=\frac{\underline{T_k^o}+\overline{T_k^o}}{2},\widetilde{Q}_{ki}=\frac{\underline{Q_{ki}}+\overline{Q_{ki}}}{2}\tag{10b}
\end{gather}
\end{subequations}
where $i\in\mathcal{R}_k, k\in\mathcal{K}_m, m\in\mathcal{M}, \phi_a$ and $\phi_b$ are nonnegative weight coefficients, $\phi_a$ represents the priority of user comfort, $\phi_b$ represents the priority of optimizing the COP. Since $\tilde{T}_{ki}^o,\tilde{Q}_{ki}$ are not controllable inputs, they are fixed by taking the average of the upper and lower bounds of the predicted values, rather than actively adjusting them to find $p_m^d$. By solving the convex optimization problem (10), the desired active power consumption of the system is given by $p_m^d=\sum_{k\in\mathcal{K}_m}\sum_{i\in\mathcal{R}_k}u_{ki}^*/(b_m-a_mT_{s_m}^*)$, where $u_{ki}^*, T_{s_m}^*$ are the optimal solution of problem (10). Clearly, $p_m^d\in[\underline{p_m},\overline{p_m}]$.


\section{Power Aggregation in Community-Level Buildings with GHP Systems}
\begin{figure}[!t]
\centering
\includegraphics[width=0.5\textwidth,trim=140 25 100 5,clip]{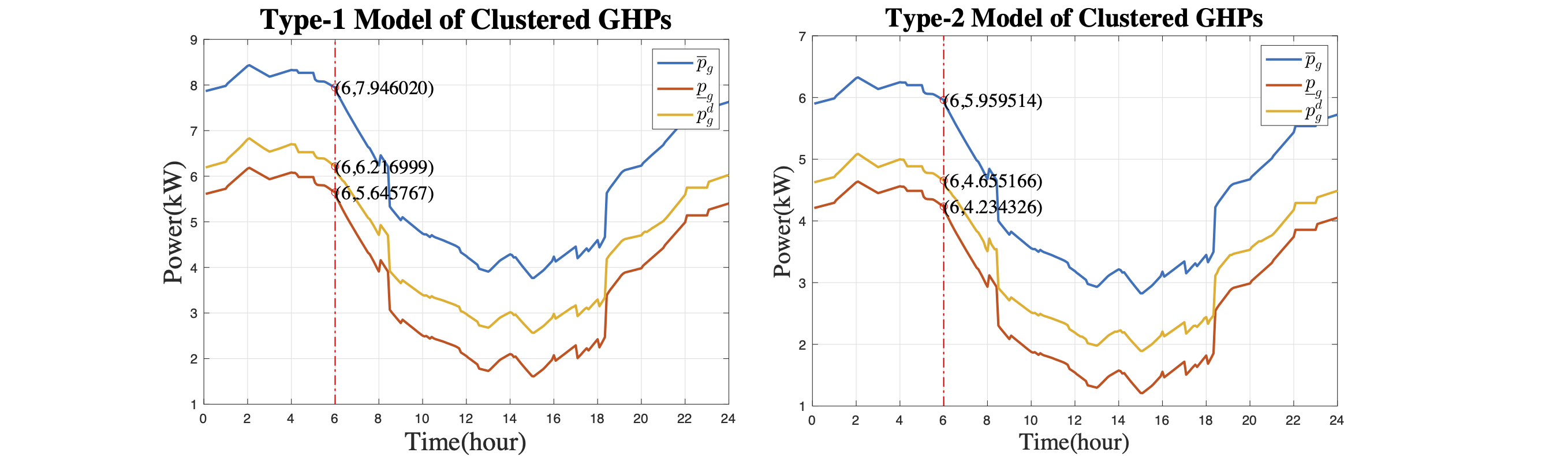}
\caption{Power consumption prediction of two clustered GHP models.}
\label{forecast}
\vspace{-0.5cm}
\end{figure}

After $[\underline{p_m},\overline{p_m}],p_m^d$ for each GHP system are obtained, the aggregator can aggregate information from clustered GHP systems in each district (i.e., each bus in the distribution network) as $[\sum_{m\in\mathcal{M}}\underline{p_m},\sum_{m\in\mathcal{M}}\overline{p_m}],\sum_{m\in\mathcal{M}}p_m^d$. In this section, we focus on the coordination and aggregation of the load resources for multiple clustered GHP systems in a distribution network. To obtain the desired optimal power at the feeder bus, a novel fully linear OPF model as~\cite{yang1} is adopted to calculate the energy demand of each GHP system to reach their desired consumption as close as possible. Then, considering the range of the energy demand of clustered GHP systems in each district (bus) under the same environment, the same OPF model is used to solve the related optimization problems to obtain the maximum and minimum values of power injection at the feeder bus. In this way, we can obtain the energy flexibility that GHP systems provide. Here, the novel OPF model, considering the linear network losses, not only retains the advantages of the linear DC OPF model but also improves the overall accuracy of power flow solutions and provides reasonable reactive and voltage magnitude estimations.


Describe the distribution network by the graph $\mathcal{G}(\mathcal{N}_0,\mathcal{L})$, where $\mathcal{N}_0$ is the set of buses, $\mathcal{N}_0:=\mathcal{N}\cup\{0\},\mathcal{N}:=\{1,2,...,N\}$, $\mathcal{L}\subset\mathcal{N}_0\times\mathcal{N}_0$ is the set of distribution lines, and bus $0$ is the feeder bus. Note that $\mathcal{M}_a$ refers to the load resource of the set of clustered GHPs at bus $a, a\in\mathcal{N}_0$. For each time granularity, we solve different optimization problems by using different utility functions in (11) but the same constraints conditions in (12) to obtain the desired, maximum and minimum values of the power injection at the feeder bus respectively:

\subsubsection{The objective functions}  
\begin{equation}\label{objectivefunc}
\min U\tag{11}
\end{equation}
where $U$ is the objective function. $U\!\!\!\!\!=\!\!\!\!\!\frac{1}{2}\sum_{a\in\mathcal{N}_0}(P_{a,{\scriptscriptstyle GHP}}\!\!-\!\!\sum_{m\in\mathcal{M}_a}\!p_m^d)^2$ is used to obtain the desired optimal active power, where $P_{a,{\scriptscriptstyle GHP}}$ is the active power demand of the clustered GHP systems at bus $a,\forall a\in\mathcal{N}_0$. $U\!=P_0^2/2$, and $U\!=-\log P_0$ are used respectively to get the minimum and maximum active power at the feeder bus. We can also consider reactive power by assuming that the fixed power coefficient of the GHP system is the constant $\eta$~\cite{factor}. 


\subsubsection{Constraints of the Linearized OPF Model}
\begin{align}\label{constraints}
&P_{ab}=g_{ab}\frac{v_a^2-v_b^2}{2}-b_{ab}\theta_{ab}+P_{ab}^L\tag{12a}\\
&Q_{ab}=-b_{ab}\frac{v_a^2-v_b^2}{2}-g_{ab}\theta_{ab}+Q_{ab}^L\tag{12b}\\
&\sum_{g\in a}P_g-\!\!P_{a,{\scriptscriptstyle GHP}}\!\!-\!\!\hat P_a\!=\!\!\sum_{(a,b)\in\mathcal{L}}\!\!P_{ab}+\!\left({\sum_{b=0}^{\mathcal{N}_0}}G_{ab}\right)\!v_a^2\tag{12c}\\
&\sum_{g\in a}Q_g-\!Q_{a,{\scriptscriptstyle GHP}}\!\!-\!\hat Q_a\!=\!\!\sum_{(a,b)\in\mathcal{L}}\!\!Q_{ab}+\!\left({\sum_{b=0}^{\mathcal{N}_0}}\!-\!B_{ab}\right)v_a^2\tag{12d}\\
&(P_{ab})^2+(Q_{ab})^2\leq S_{ab,\max}^2,(a,b)\in\mathcal{L}\tag{12e}\\
&P_g^{\min}\leq P_g\leq P_g^{\max}, Q_g^{\min}\leq Q_g\leq Q_g^{\max}\tag{12f}\\
&\left(v_{a}^{\min}\right)^2\leq v_a^2\leq\left(v_a^{\max}\right)^2, a\in\mathcal{N}_0\tag{12g}\\
&\sum_{m\in\mathcal{M}_a}\underline{p_m}\leq P_{a,{\scriptscriptstyle GHP}}\leq\sum_{m\in\mathcal{M}_a}\overline{p_m}, a\in\mathcal{N}_0\tag{12h}
\end{align}where $(a,b)$ refers to the branch, $P_{\{\cdot\}}/Q_{\{\cdot\}}$ is the active/reactive power, $P_{\{\cdot\}}^L/Q_{\{\cdot\}}^L$ is the active/reactive power loss, $P_g^{\{\cdot\}}/Q_g^{\{\cdot\}}$ is the active/reactive power of the generator, $G_{ab}/B_{ab}$ is the real/imaginary part of the entry of the admittance matrix, $g_{ab}/b_{ab}$ is the conductance/susceptance of branch $(a,b)$, $S_{ab,max}$ is the apparent power limitation, $Q_{a,{\scriptscriptstyle GHP}}$ is the reactive power demand of the clustered GHP systems at bus $a,\forall a\in\mathcal{N}_0$, $\hat P_a/\hat Q_a$ is the active/reactive power demand of the fixed load at bus $a,\forall a\in\mathcal{N}_0$. Equations $(12a)-(12b)$ are the power flow equations, $(12c)-(12d)$ are the nodal power balance equations, $(12e)$ is the branch flow limits, $(12f)-(12h)$ are the operational constraints (detailed modeling of the linearized OPF model is presented in~\cite{yang1}). 


\section{A Numerical Example}
As shown in Figure~\ref{der}, an IEEE $33$-bus distribution network model is used in the case study for numerical demonstrations. We set up an example of aggregating $18$ GHP systems, including two types of clustered GHP systems. The first model contains three GHP systems, providing heating to three identical buildings with radiators respectively, each building with four heat zones. The second model provides heating to two identical buildings respectively with radiators through two GHP systems (for each GHP system, we do not consider more heat zones because our approach is scalable). We consider adding the first clustered GHP model at bus 5,6,25, and the second clustered GHP model at bus 1,2,18,22 (the fixed power factor of the GHP system is set as $0.95$), respectively. 

\begin{figure}[!t]
\centering
\includegraphics[width=0.5\textwidth,trim=15 5 20 5,clip]{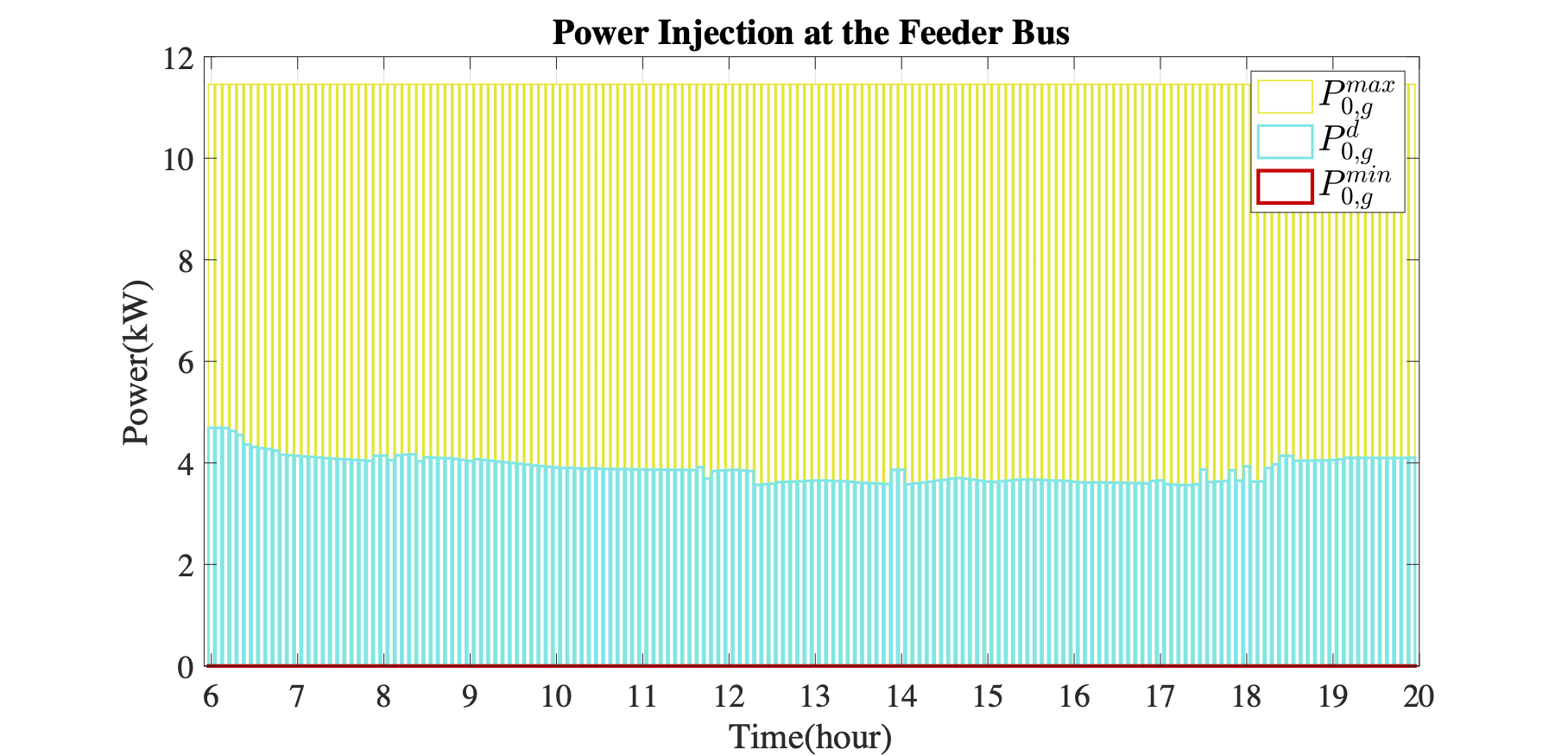}
\caption{The desired aggregated power and the optimal range via the fully linear OPF model.}
\label{pc}
\vspace{-0.5cm}
\end{figure}

In the forecast procedure, set $5$ minutes as the time granularity, and we assume that (i) the prediction accuracy of disturbances is within $\pm2^{\circ}C$ and $\pm20\%$ of their real values, (ii) the prediction of the outdoor temperature is the same for all buildings. The simulation parameters are derived from \cite{DATA}. Figure~\ref{forecast} gives the forecast result of short-term energy demand for the two types of clustered GHP systems. We also use $5$ minutes as the time granularity during the aggregation process. We use the CVX package in Matlab R2020b to solve the proposed optimization model in Section III. As illustrated in Figure \ref{pc}, the desired active power injection and the range of active power injection from 6:00 to 20:00 are displayed. In fact, the power injection value varies at different moments, but because the power value of GHP inputs $P_{a,GHP}$ are small, it has little influence on the power injection at the feeder bus. The yellow and blue parts in Figure \ref{pc} are the energy flexibility that GHP systems provide. The results show that the study on the integration and operation framework of GHP systems in the distribution network is meaningful and effective since GHP systems could provide energy flexibility to the distribution network. 

\section{Conclusion and Future Work}
In this paper, we propose an integration and operation framework for multiple clustered GHPs in the distribution network, which uses a hierarchical communication and optimization method to coordinate all GHPs. In the future, we will further consider a decomposition framework of aggregated power and apply the OPF model to determine the power allocation decision at each bus. Then, we will apply the whole integration and decomposition framework to the heat distribution subsystem, together with floor heating and radiators.

\vspace{12pt}
\end{document}